\documentstyle [12pt,epsf]{article}

\input{epsf}
\begin{document}
\begin{titlepage}
\begin{center}

\vspace{0.4in}

{\Large \bf The Effective Potential for Composite Operator in the
Scalar Model at Finite Temperature}\\
\vspace{.3in}
{\large\em G.N.J.A\~na\~nos$ ^{1,2}$ and N.F.Svaiter$ ^1$\\
$^1$Centro Brasileiro de Pesquisas F\'{\i}sicas-CBPF\\ Rua
Dr.Xavier
 Sigaud 150, Rio de Janeiro, RJ 22290-180 Brazil\\ E-mail: nfuxsvai@lafex.cbpf.br \\
$^2$Laborat\'orio Nacional de Computa\c{c}\~ao Cient\'{\i}fica-LNCC\\
 Av. Get\'ulio Vargas, 333 - Quitandinha - Petr\'opolis, RJ 25651-070
 Brazil\\ E-mail: gino@lafex.cbpf.br}
\end{center}
\subsection*{Abstract}

We discuss the  $\varphi^4$ and $\varphi^6 $ theory defined in a flat $D$-dimensional space-time. We assume that the system is in equilibrium with a thermal bath at temperature $\beta^{-1}$. To obtain non-perturbative result, the $ 1/N $ expansion is used. The method of the composite operator (CJT) for summing a large set of Feynman graphs, is developed for the finite temperature
system. The ressumed effective potential and the analysis of the $D=3$ and
$D=4$ cases are given.


\end{titlepage}

\newpage

\baselineskip .37in
\section {Introduction}

The conventional perturbation theory in the coupling constant or in $\hbar$ 
i.e., the loop expansion can only be used for the study of small quantum
corrections to classical results. When discussing quantum
mechanical effects to any given order in such an expansion, one
is not usually able to justify the neglect of yet higher
 order. In other words, for theories with a large $N$ dimensional internal
symmetry group, there exist another perturbation scheme, the $1/N$ expansion,
which circumvents this criticism. Each term in the $ 1/N $ expansion
contains an infinite subset of terms of the loop expansion.
The $ 1/N $ expansion has the nice property that the leading-order quantum
 corrections are of the same order as the classical quantities. Consequently, the leading
 order which adequately characterizes the theory in the large $N$ limit preserves much of
 the nonlinear structure of the full theory.
In the next section we derive the effective action to leading
order in $1/N$ in $D$-dimensional space-time and consequently the effective
potential. It is known that, in $D>4$, such theories with
$\varphi^4$ interaction are in fact free field theory, while in
$D<4$ they have a non-trivial continuum limit  as an interacting
field theory. For $D=3$ it has been shown that, in the large $N$ limit, the
 $\varphi^6$ theory has a UV fixed point and therefore must have a second IR
 fixed point \cite{bardeen}. At least for large $N$  the $(\varphi^6)_{D=3}$ theory is
 known to be qualitatively different from $(\varphi^4)_{D=4}$ theory.

In a previous work \cite{ananos} by use of the composite operator formalism
and, we re-examinate the behavior at
finite temperature of the $O(N)$-symmetric $\lambda\varphi^{4}$
model in a generic D-dimensional Euclidean space. In the cases
$D=3$ and $D=4$, an analysis of the thermal behavior of the
renormalized squared mass and coupling constant are done for all
temperatures. It results that the thermal renormalized squared
mass is positive and increases monotonically with the
temperature. It is interesting to stress that the behavior of the thermal coupling
constant is quite different in odd or even dimensional space. In $D=3$, the thermal
coupling constant decreases up to a minimum value
different from zero and then grows up monotonically as the
temperature increases. In the case $D=4$, it is found that the
thermal renormalized coupling constant tends in the high
temperature limit to a constant asymptotic value. Also for
general D-dimensional Euclidean space, we are able to obtain a
formula for the critical temperature of the second order phase
transition. This formula agrees with previous known results at
$D=3$ and $D=4$ \cite{jon,bimo}.

It is well known that the introduction of the $\varphi^6$ term
generated a rich phase diagram, with the possibility of second
order, first order phase transitions or even both transitions
occurring simultaneously. This situation defines the tricritical
phenomenon. Some systems such antiferromagnets in the presence of
a strong external field or the $He^3-He^4$ mixture exhibits such
behavior. In a previous paper the massive $(\varphi^6)_{D=3}$
model was analyzed at finite temperature at the two-loops
approximation. We demonstrate the existence of the tricritical
point \cite{an}. A natural extension of this paper was done in
ref. \cite{pre}. In this paper we proved the existence of the
tricritical point using a non-pertubative approach. This was done
using the CJT formalism i.e. the composite operator formalism
\cite{cornwall}.

Here we continuous to study the composite operator method. Still studying the  $\varphi^6$
theory in the large N expansion, the effective potential at finite temperature is calculated.
The organization of the
paper is the following. In section II we derive the effective
potential using the composite operator (CJT) formalism. In section
III the thermal effective potential is found for a D-dimensional generic
space.
Conclusions are given in section IV. In this paper we use
$\hbar=c=k_{b}=1$.

\section{The effective potential (The CJT formalism)}

We are interested here in the most general renormalizable scalar
field model $\lambda \varphi^{4}+\eta\varphi^{6}$ possessing an
internal symmetry $ O(N)$, in a generic $D$-dimensional space-time. Of course, for $D=4$
this theory is non-renormalizable. In this case to ensure renomalisability we must make
$\eta=0$. Let us define at the beginning the field in a generic D-dimensional  space-time.
For simplicity we will call this theory a $\varphi^{6}$ model.

Using the method of composite operator developed by Cornwall,
Jackiw and Tomboulis \cite{cornwall,livro}, Townsend derived the
effective potential of $\varphi^6$ theory in the $1/N$ expansion
for $D=3$ at zero temperature \cite{townsend}. This author proved that $1/N$ expansion is consistent for $\varphi^6$ to leading order.

The Lagrangian
density of the $O(N)$ symmetry  $\varphi^6$ theory is :
\begin{equation}
{\cal L}(\varphi)=\frac{1}{2} (\partial_{\mu} \varphi)^2-\frac{1}{2}m_{0}^2
\varphi^2-\frac{\lambda_0}{4!N} \varphi^4- \frac{\eta_0}{6! N^2}\varphi^6,
\end{equation}
where the quantum field is an $N$-component vector
in the $N$-dimensional
internal symmetry space.
For definiteness, we work at zero-temperature; however,
the finite temperature
generalizations can be easily obtained \cite{dolan}.
We are interested in the effective action $\Gamma (\phi)$ which governs
the behavior of the expectation values $\varphi_a (x)$ of the quantum field
where $\phi$ is given by
\begin{equation}
\phi(x) \equiv {\delta W(J) \over \delta J(x)} = <0|\varphi(x)|0>,
\end{equation}
where $W(J)$ is the generating functional for connected Green's functions.

$\Gamma (\phi)$ can be shown to be the sum of one-particle irreducible (1PI)
Feynman graphs with a factor $\phi_a (x)$ on the external line.
We make use of the formalism of composite operator which reduces  the problem
 to summing two particle irreducible (2PI) Feynman graphs by defining a
generalized effective action $\Gamma (\phi,G)$ which is a
functional not only of $\phi_a (x)$, but also of the expectation
values $G_{ab} (x,y)$ of the time ordered product of quantum
fields $<0|T(\varphi(x)\varphi(y))|0> $, i.e.
\begin{equation}
 \Gamma (\phi,G)= I(\phi)+\frac{i}{2} TrLn G^{-1} +
\frac{i}{2} Tr D^{-1}(\phi) G + \Gamma_2(\phi,G) +\dots \;\; ,
\label{G1}
\end{equation}
where $I(\phi)=\int dx^D {\cal L}(\phi) $, $G$ and $D$ are
matrices in both the functional and the internal space whose
elements  are $G_{ab}(x,y)$, $D_{ab}(\phi;x,y)$ respectively and
$D$ is defined by
\begin{equation}
i D^{-1}=\frac{\delta^2 I(\phi)}{\delta\phi(x) \delta\phi(y)}.
 \end{equation}
The quantity $\Gamma_2(\phi,G)$ is computed as
follows. In the classical action $I(\varphi)$ we have to shift the field $\varphi$ by
$\phi$. The new action  $I(\varphi+\phi) $  possesses terms cubic and higher in
$\varphi$. This define an interaction part   $I_{int} (\varphi,\phi)$
where the vertices depend on $\phi$.  $\Gamma_2(\phi,G)$ is given by
sum of all  (2PI) vacuum graphs  in a theory with vertices determined by $I_{int} (\varphi,\phi)$ and the propagators set equal to
$G(x,y)$. The trace and logarithm in eq.(\ref{G1}) are functional. After these
procedures the interaction Lagrangian density becomes
\begin{eqnarray}
 {\cal L}_{int}(\varphi,\phi)&=&-\frac{1}{2} \left( \frac{\lambda_0 \phi_a}{3N} + \frac{\eta_0 \phi^2\phi_a}{30N^2} \right) \varphi_a \varphi^2 -
 \left( \frac{8\eta_0 \phi_a\phi_b\phi_c}{6N^2} \right) \varphi_a \varphi_b
\varphi_c -\frac{1}{4!N} \left( \lambda_0+ \frac{\eta_0\phi^2}{10N}\right)
 \varphi^4 \nonumber \\
& &
-\left( \frac{12\eta_0\phi_a\phi_b}{6!N^2} \right)\varphi_a\varphi_b\varphi^2
-\frac{1}{5!}\left( \frac{\eta_0\phi_a}{N^2}\varphi_a\varphi^4\right)
-\frac{\eta_0}{6!N^2}\varphi^6.
\end{eqnarray}
The effective action $\Gamma(\phi)$ is found by solving for $G_{ab}(x,y)$  the equation
\begin{equation}
\frac{\delta \Gamma(\phi,G)}{\delta G_{ab}(x,y)}=0,
\label{eq1}
\end{equation}
and substituting the solution in the generalized effective action $\Gamma(\phi,G)$.

The vertices in the above equation contain factors of $1/N$ or $1/N^2$, but
a $\varphi$ loop gives a factor of N provided the $O(N)$ isospin flows
around it alone and not into another part of the graph. We usually call
such loops bubbles. Then at leading order in $1/N$, the vacuum graphs are
bubble trees with two or three bubbles at each vertex. The (2PI) graphs
are shown in figure.(1).
\begin{figure}[ht]

\centerline{\epsfysize=1.0in\epsffile{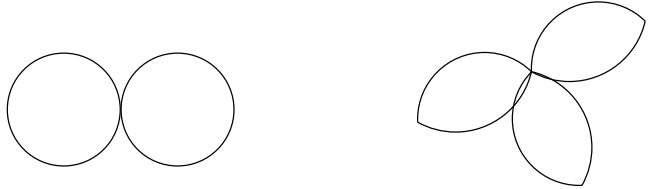}}
\caption[region]
{The 2PI vacuum graphs }
\end{figure}
It is straightforward to obtain
\begin{equation}
\Gamma_2(\phi,G)=\frac{-1}{4!N}\int d^{D}x \left( \lambda_0+
\frac{\eta_0\phi^2}{10N} \right) [G_{aa}(x,x)]^2 -
\frac{\eta_0}{6!N^2}
\int d^{D}x [G_{aa}(x,x)]^3 .
\label{eq2}
\end{equation}

Therefore eq.(\ref{eq1}) becomes

\begin{eqnarray}
\frac{\delta\Gamma(\phi,G)}{\delta G_{ab}(x,y)}&=& \frac{1}{2}(G^{-1})_{ab}
(x,y) + \frac{i}{2} D^{-1}(\phi)-\frac{1}{12N}\left( \lambda_0+
\frac{\eta_0\phi^2}{10N}\right)[\delta_{ab}G_{cc}(x,x)]\delta^D(x-y)
\nonumber \\
& & -\frac{3\eta_0}{6!N} \delta_{ab}[G_{cc}(x,x)]^2\delta^D(x-y)=0.
\label{eq3}
\end{eqnarray}
Rewriting this equation, we obtain the gap equation
 \begin{eqnarray}
(G^{-1})_{ab}(x,y) &=& D^{-1}_{ab}(\phi;x,y)+ \frac{i }{6N}\left( \lambda_0+
\frac{\eta_0\phi^2}{10N}\right)[\delta_{ab}G_{cc}(x,x)]\delta^D(x-y)+\nonumber
\\
 & &\frac{i\eta_0}{5!N^2} \delta_{ab}[G_{cc}(x,x)]^2\delta^D(x-y).
\label{eq4}
\end{eqnarray}
Hence
\begin{equation}
\frac{i}{2} Tr D^{-1}G= \frac{1}{12N}\int d^Dx \left( \lambda_0+\frac{\eta_0\phi^2}{10N} \right) [G_{aa}(x,x)]^2 +
\frac{3\eta_0}{6!N^2}\int d^{D}x [G_{aa}(x,x)]^3+ cte .
\label{eq5}
\end{equation}
Using eqs.(\ref{eq4}) and eq.(\ref{eq5}) in eq.(\ref{eq2}) we find the effective action
\begin{eqnarray}
\Gamma(\phi) &=& I(\phi)+\frac{i}{2} Tr[Ln G^{-1}]+\frac{1}{4!N}
\int d^Dx \left( \lambda_0+\frac{\eta_0\phi^2}{10N} \right) [G_{aa}(x,x)]^2+
 \nonumber \\
& & \frac{2\eta_0}{6!N^2} \int d^Dx [G_{aa}(x,x)]^3 ,
\label{effa}
\end{eqnarray}
where $G_{ab}$ is given implicitly by eq.(\ref{eq4}). The trace in (\ref{effa}) are both the functional and the internal space.
The last two terms on the r.h.s of eq.(\ref{effa}) are the leading contribution to the effective action in the $1/N$ expansion.
As usual we may simplify the situation by separating $G_{ab}$ into transverse and longitudinal components, so
\begin{equation}
G_{ab}=(\delta_{ab}-\frac{\phi_a \phi_b}{\phi^2})g+\frac{\phi_a \phi_b}{\phi^2} \stackrel{\sim}{g}\; ,
\end{equation}
in this form we can invert $G_{ab}$,
\begin{equation}
(G)^{-1}_{ab}=(\delta_{ab}-\frac{\phi_a \phi_b }{\phi^2} ) g^{-1} +
\frac{\phi_a \phi_b}{\phi^2} {\stackrel{\sim}{g}}^{-1}\; .
\end{equation}

Now we can take the trace with respect to the indices of the internal space,
\begin{equation}
G_{aa}=Ng +O(1) ,\;\;\;\; (G)^{-1}_{aa}=Ng^{-1}+ O(1) \; .
\label{gd}
\end{equation}
From this equation at leading order in $1/N$, $G_{ab}$ is diagonal in $a,b$. Substituting
eq.(\ref{gd}) into eq.(\ref{effa}) and
eq.(\ref{eq4}) and keeping only the leading order one finds that the daisy and
superdaisy resummed effective potential for the $\varphi^6$ theory is given by:
\begin{eqnarray}
\Gamma(\phi)&=&I(\phi)+\frac{iN}{2} tr(\ln g^{-1})+\frac{N}{4!}
\int d^Dx \left( \lambda_0+\frac{\eta_0\phi^2}{10N} \right) g^2(x,x)+
\nonumber \\
 & & \frac{2N\eta_0}{6!} \int d^Dx g^3(x,x) +O(1) ,
\label{effa1}
\end{eqnarray}
where the trace is only in the functional space, and the gap equation becomes
\begin{equation}
g^{-1}(x,y)=i\left[ \Box + m_0^2+
\frac{\lambda_0}{6}(\frac{\phi^2}{N}+ g(x,x)) + \frac{\eta_0}{5!}
(\frac{\phi^2}{N}+ g(x,x))^2 \right ] \delta^{D}(x-y)+ O(\frac{1}{N})
\label{gap1}.
\end{equation}
It is important to point out that this calculation was done by Townsend \cite{townsend}. We
interested to generalize these results assuming that the
system is in equilibrium with a thermal bath a temperature $T=\beta^{-1}$.  Since we are
studying the equilibrium situation it is convenience to use the Matsubara formalism.
Consequently it is convenient to continue all momenta to Euclidean values $(p_0=ip_4)$
and take the following Ansatz for $g(x,y)$,
\begin{equation}
g(x,y)=\int \frac{d^{D}p}{(2\pi)^D}\frac{\exp^{i(x-y)p}}{p^2+M^2(\phi)}.
\label{ge}
\end{equation}
Substituting eq.(\ref{ge}) in eq.(\ref{gap1}) we get the expression for
the gap equation:
\begin{equation}
 M^2(\phi)=m^2_0+
\frac{\lambda_0}{6}\left( \frac{\phi^2}{N}+F(\phi)\right) + \frac{\eta_0}{5!}
\left( \frac{\phi^2}{N}+ F(\phi)\right)^2,
\label{M2}
\end{equation}
where $F(\phi)$ is given by
\begin{equation}
F(\phi)=\int \frac{d^Dp}{(2\pi)^D}\frac{1}{p^2+M^2(\phi)},
\label{Fphi}
\end{equation}
and the effective potential in the D-dimensional Euclidean space can be expressed as
\begin{equation}
V(\phi)=V_0(\phi)+\frac{N}{2}\int \frac{d^Dp}{(2\pi)^D} \ln \left
[p^2+M^2(\phi) \right ]
 -\frac{N}{4!}(\lambda_0+\frac{\eta_0 \phi^2}{10N}) F(\phi)^2-
\frac{2N\eta_0 F(\phi)^3}{6!} ,
\label{ve1}
\end{equation}
where $V_0(\phi)$ is the classical potential.
In the next section using the Matsubara formalism we present the effective potential of the model at finite temperature.

\section{The effective potential for $\varphi^6$ theory
 at finite temperature}

Let us suppose that our system is in equilibrium with a thermal
bath. To study the temperature effects in quantum field theory we will use the imaginary time Green function approach \cite{dolan}. In this formalism the Euclidean-time $\tau$ is restricted to the interval $0 \leq \tau \leq \beta$, and the bosonic filed satisfies periodic boundary conditions in Euclidean-time.
This is equivalence to replace the continuous four momenta $k_4$ by discrete
$\omega_n$ and the integration by a summation
($\beta=\frac{1}{T}$):
\begin{eqnarray}
 k_4 &\rightarrow& \omega_n =\frac{2\pi n}{\beta}, \;\;\;\;\; n=0,\pm1,\pm2,...
\nonumber \\
\int\frac{d^Dk}{(2\pi)^D} &\rightarrow &  \sum_{n} \frac{1}{\beta}\;\;
\int\frac{d^{D-1}k}{(2\pi)^{D-1}} .
\end{eqnarray}
It is important to stress that all the Feynman rule are the same as the temperature case, except, as we stressed that momentum space integrals over the zeroth component is replace by summ over discret summs.
The effective potential at finite temperature can be write as:
\begin{eqnarray}
V_{\beta}(\phi)&= &
V_0(\phi)+\frac{N}{2\beta}\sum_{n}^{\infty} \int
\frac{d^{D-1}p}{(2\pi)^{D-1}} \ln \left
[\omega_n+p^2+M_{\beta}^2(\phi) \right ] -
\nonumber \\
& &\frac{N}{4!}(\lambda_0+\frac{\eta_0 \phi^2}{10N})
F_{\beta}(\phi)^2-\frac{2N\eta F_{\beta}(\phi)^3}{6!} .
\label{ve3}
\end{eqnarray}
where $F_{\beta}(\phi)$ is a finite temperature generalization of
$F(\phi)$, where,
\begin{equation}
F_{\beta}(\phi)=\frac{1}{\beta}\sum_{n=-\infty}^{\infty} \int
\frac{d^{D-1}p}{(2\pi)^{D-1}}
\frac{1}{\omega^2_n+p^2+M^2_{\beta}(\phi)}\; . \label{Fbphi}
\end{equation}
The gap equation for this theory at finite temperature is
given by,
\begin{equation}
 M^2_{\beta}(\phi)=m^2_0+
\frac{\lambda_0}{6}\left( \frac{\phi^2}{N}+F_{\beta}(\phi)\right)
+ \frac{\eta_0}{5!} \left( \frac{\phi^2}{N}+
F_{\beta}(\phi)\right)^2. \label{M2T1}
\end{equation}
 In order to regularize
$F_{\beta}(\phi)$ given by eq.(\ref{Fbphi}), we use a mixing
between dimensional regularization and analytic regularization.
For this purpose we define the expression $I_{\beta}(D,s,m)$ as :
 \begin{equation}
I_{\beta}(D,s,m)=\frac{1}{\beta}\sum_{n=-\infty}^{\infty}\int
\frac{d^{D-1}k}{(2\pi)^{D-1}}\frac{1}{(\omega^2_n+k^2+m^2)^s} \; .
\end{equation}
Using the analytic extension of the inhomogeneous Epstein zeta
function it is possible to obtain the analytic extension of
$I_{\beta}(D,s,m)$;
\begin{equation}
I_{\beta}(D,s,m)= \frac{m^{D-2s}}{(2\pi^{1/2})^D\Gamma(s)} \left[
\Gamma(s-\frac{D}{2}) + 4 \sum_{n=1}^{\infty} \left(
\frac{2}{mn\beta}
 \right)^{D/2-s} K_{D/2-s}(mn\beta) \right]
\end{equation}
where $K_{\mu}(z)$ is the modified  Bessel function of the third
kind. Fortunately for $D=3$ the analytic extension of the
function $I_{\beta}(D,s=1,m=M_\beta(\phi))=F_{\beta}(\phi)$ is
finite and can be expressed in a closed form \cite{an} (note that
in $D=3$ we have no pole, at least in this approximation), and in
particular as
\begin{equation}
F_{\beta}(\phi)=I_{\beta}(3,1,M_{\beta}(\phi))=-\frac{M_{\beta}(\phi)}{4\pi}\left(
1+\frac{2\ln(1-e^{-M_{\beta}(\phi)\beta})}{M_{\beta}(\phi)\beta}
\right ). \label{m2b2}
\end{equation}
This result is not a peculiarity of this method of
regularization, because this happens also in dimensional
regularization at zero temperature, that is, in odd dimensions
integrals which are divergent by naive power counting may be to
regulated to finite value with no poles occurring, for example in
$D=3$. In order to regularize the second term of eq.(\ref{ve3}),
we use the following method: We define,
\begin{equation}
LF_\beta(\phi)=\frac{1}{\beta}\sum_{n=1}^{\infty} \int
\frac{d^{D-1}p}{(2\pi)^{D-1}} \ln \left
[\omega_n+p^2+M_{\beta}^2(\phi) \right ]
\end{equation}
then,

 \begin{equation}
\frac{ \partial LF_\beta(\phi)}{\partial
M_{\beta}}=(2M_{\beta})\frac{1}{\beta}\sum_{n=1}^{\infty} \int
\frac{d^{D-1}p}{(2\pi)^{D-1}}
\frac{1}{\omega_n+p^2+M_{\beta}^2(\phi)}
\end{equation}
and from the equation (\ref{Fbphi}), we have that,
 \begin{equation}\label{lfb}
\frac{ \partial LF_\beta(\phi)}{\partial M_{\beta}}=(2M_{\beta})
F_{\beta}(\phi),
\end{equation}
in this way the effective potential could be  regularized. For
 $D=3$, $F_{\beta}(\phi)$ is finite and is given by eq. (\ref{m2b2})\cite{an}
  and integrating the eq.(\ref{lfb}), we obtain:
\begin{equation}\label{m2b2}
LF_\beta(\phi)_R=-\,{\frac {{M_{\beta}(\phi)}^{3}}{6\pi }}-{\frac
{M_{\beta}(\phi) Li_2({e^{-M_{\beta}(\phi)\beta}})} {\pi
\,{\beta}^{2}}}-{\frac {Li_3({e^{-M_{\beta}(\phi)\beta}})}{\pi
\,{\beta}^{3}}} .
\end{equation}

The definition of general polylogarithm function $Li_n(z)$ can be found in ref. \cite{poly}.

The daisy and super daisy resummed effective potential at
finite temperature for $D=3$ is given by:
\begin{equation}
V_{\beta}(\phi)=V_0(\phi)+\frac{N}{2} LF_\beta(\phi)_R-
\frac{N}{4!}(\lambda_0+\frac{\eta_0 \phi^2}{10N})
(F_{\beta}(\phi)_R)^2 -\frac{2N\eta (F_{\beta}(\phi)_R)^3}{6!} .
\label{ve4}
\end{equation}
and the gap equation (see eq.(\ref{M2T1})):
\begin{eqnarray}\label{m2d3}
 M^2_{\beta}(\phi) & = &  m^2_0+
\frac{\lambda_0}{6}\left(
\frac{\phi^2}{N}-\frac{M_{\beta}(\phi)}{4\pi} \left[
1+\frac{2\ln(1-e^{-M_{\beta}(\phi)\beta})}{M_{\beta}(\phi)\beta}
\right ]\right)  \nonumber \\
&& + \frac{\eta_0}{5!} \left(
\frac{\phi^2}{N}-\frac{M_{\beta}(\phi)}{4\pi} \left[
1+\frac{2\ln(1-e^{-M_{\beta}(\phi)\beta})}{M_{\beta}(\phi)\beta}
\right]\right)^2 .
\end{eqnarray}
For the case $D=4$, and for  $\eta=0$, where the theory is just
renormalizable, the effective potential can be obtain in the same
way, that is:
\begin{equation}
V_{\beta}(\phi)=V_0(\phi)+\frac{N}{2} (LF_\beta(\phi)_R)_{D=4}-
\frac{\lambda_0 N}{4!}(F_{\beta}(\phi)_R)_{D=4}^2 , \label{ve31}
\end{equation}
where $V_0$ is the classical potential,
\begin{equation}
V_0(\phi)=\frac{1}{2}m^2 \varphi^2+\frac{\lambda}{4!N} \varphi^4,
\end{equation}
and $(F_{\beta}(\phi)_R)_{D=4}$ is given by:
\begin{equation}
(F_{\beta}(\phi)_R)_{D=4}=\frac{\lambda M_{\beta}^2(\phi)}{2\pi
^2}\int_1^\infty \frac{(p^2-1)^{\frac
12}}{e^{M_{\beta}(\phi)\beta p}-1}dp \quad .
\end{equation}
and in the limit of high temperature, we could write the above
equation as:
\begin{equation}
(F_{\beta}(\phi)_R)_{D=4}= \frac{1}{12\beta^2}-\frac {M_{\beta}(\phi)}{4\pi \beta }-\frac{M_{\beta}^2(\phi)}{8\pi^2}
\ln {(M_{\beta}(\phi)\beta)};
\end{equation} and,
\begin{equation}
(LF_{\beta}(\phi)_R)_{D=4} = {\frac
{{M_{\beta}(\phi)}^{2}}{12{\beta}^{2}}}-
{\frac{{M_{\beta}(\phi)}^{3}}{6\pi\beta}}- {\frac
{{M_{\beta}(\phi)}^{4}\ln (M_{\beta}(\phi)\beta)}{16{\pi}^{2}}}+
 {\frac {{M_{\beta}(\phi)}^{4}}{64{\pi}^{2}}},
\end{equation}
and the gap equation for $D=4$ is given by:
\begin{equation}
 M^2_{\beta}(\phi)=\tilde{m}^2(\phi)+
\frac{\lambda}{6}\left( \frac{1}{12\beta^2}-\frac {M_{\beta}(\phi)}{4\pi \beta }-\frac{M_{\beta}^2(\phi)}{8\pi ^2}
\ln {(M_{\beta}(\phi)\beta )}\right),
\label{M2T2}
\end{equation}
where $\tilde{m}^2(\phi)=m^2+\frac{\lambda}{6}\frac{\phi^2}{N}$,
and $m^2$, $\lambda$ are the renormalized  mass and coupling
constant at zero temperature respectively. We note that, from the
gap equation in eq.(\ref{M2T2}), we find that for the coupling
constant $\lambda \ll 1$, the condition
$M^2_{\beta}(\phi)/T^2\ll1$ is  consistent with
$\tilde{m}^2(\phi)/T^2$, which is exactly the required condition
 for the high temperature expansion \cite{dolan}.
\section{Conclusions}
In this paper we have performed an analysis of the daisy and
super daisy effective potential for the theory $\varphi^{4}$ and
 $\varphi^6$ in  $D$-dimensional Euclidean space at finite temperature.
The form of effective potential have been found explicitly using
resummation method in the leading order $1/N$ approximation
(Hartree-Fock approximation). We have seen how dimensional
regularization and analytic regularization  can be used to compute
the effective potential at finite temperature. In odd dimensional
theory when power counting indicates that the diverges should
occur, dimensional regularization and analytic  does not give
rise to a pole.

\section{Acknowlegements}

This paper was supported by Conselho Nacional de
Desenvolvimento Cientifico e Tecnologico do Brazil (CNPq).

\end{document}